\documentclass[Journal]{IEEEtran}
\IEEEoverridecommandlockouts

\usepackage{cite}
\usepackage{amsmath,amssymb,amsfonts}
\usepackage{algorithmic}
\usepackage{graphicx}
\usepackage{textcomp}

\usepackage{csvsimple}
\usepackage{subfig}
\usepackage{breqn}
\usepackage{enumerate}
\usepackage{algorithm2e}
\usepackage{diagbox}
\usepackage{pict2e}
\usepackage{longtable}
\usepackage{float}
\usepackage{mathtools}

\usepackage[english]{babel}
\usepackage{blindtext}
\usepackage{stfloats}

\usepackage{caption}
\makeatletter
\def\ps@IEEEtitlepagestyle{ 
    \def\@oddfoot{\mycopyrightnotice}
}
\def\mycopyrightnotice{
    {978-1-7281-8480-7/21/\$31.00~\copyright~2021 IEEE\hfill}
}

\begin{document}
	
	\graphicspath{{Figure/}}

\title{Quantized-but-uncoded Distributed Detection (QDD) with Unreliable Reporting Channels}

\author{Lei Cao, {\it Senior Member, IEEE} and Ramanarayanan Viswanathan, {\it Life Fellow, IEEE}\\
\thanks{L. Cao and R. Viswanathan are with Department of Electrical and Computer Engineering, The University of Mississippi, University, MS 38677 USA (e-mail: lcao@olemiss.edu; viswa@olemiss.edu). This paper was presented in part at \cite{cao23:distributed}}
 }

\maketitle


\begin{abstract}
Distributed detection primarily centers around two approaches: Unquantized Distributed Detection (UDD), where each sensor reports its complete observation to the fusion center (FC),  and quantized-and-Coded DD (CDD), where each sensor first partitions the observation space and then reports to the FC a codeword. In this paper, we introduce Quantized-but-uncoded DD (QDD), where each sensor, after quantization, transmits a summarized value, instead of a codeword, to the FC. We show that QDD well adapts to the constraint of transmission power when compared to CDD, albeit with increased complexity in parameter selection. Moreover, we establish that, in the presence of independent observations, QDD upholds a necessary condition inherent in CDD. Specifically, the optimal sensor decision rules are the likelihood ratio quantizers (LRQ), irrelevant to the channel conditions. In the context of a single-sensor scenario involving binary decision at the sensor, we find that the optimal sensor rule in QDD is in general no longer ``channel blind", a feature presented in CDD. In addition, we compare these systems numerically under the same transmission power and bandwidth, while assuming additive white Gaussian noise (AWGN) in both sensing and reporting stages. Finally, we present some potential directions for future research. 
	
\end{abstract}

\begin{IEEEkeywords}
	Distributed detection, centralized or decentralized, quantize or not quantize, code or not code. 
\end{IEEEkeywords}

\section{Introduction}

\begin{figure}[htb]
\centering
    \includegraphics[scale=0.30]{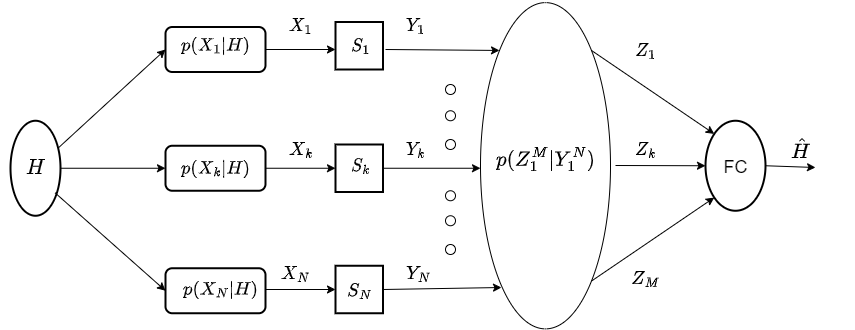}
      \caption{Sensor networks with parallel topology}
  \label{diagram}
\end{figure}

We consider deciding a binary hypothesis, with $s=-\mu$ under $H_0$ and $s=\mu$ under $H_1$, respectively, in a parallel sensing structure with $N$ sensors. The prior probabilities of hypotheses are $\pi_0$ and $\pi_1$, where $\pi_0+ \pi_1=1$.  As shown in Fig. \ref{diagram}, sensing at different sensors is independent through a noisy environment denoted as $p(X_k|H_j), j=0, 1$. Each sensor processes its sensed information $X_k$ and sends $Y_k=\gamma_k(X_k)$, either as the raw data possibly with a scaling factor or as a summarized representation, to the fusion center (FC). $\{Y_1, \dots, Y_N\} $ are reported and arrive at the FC as $\{Z_1, \dots, Z_M\}$ via a conditional probability distribution $p(Z_1, \dots, Z_M|Y_1, \dots, Y_N)$. When $M=N$ and reporting channels are independent, $p(Z_1, \dots, Z_M|Y_1, \dots, Y_N)=\Pi_{k=1}^N p(Z_k|Y_k)$.  Based on the received information, FC then makes a final decision $\hat{H}=\gamma_0(Z_1, \dots, Z_N)$, in favor of either $ H_0$ or $H_1$. The Bayes' error is generally used as the performance metric in this work, i.e., $P_e= P_f \pi_0+P_m\pi_1$, where $P_f=P(\hat{H}=H_1|H_0)$ and $P_m=P(\hat{H}=H_0|H_1)$ are the false alarm and miss-detection probabilities, respectively. Since the optimal decision rule in the FC is always the likelihood ratio test (LRT), the objective of a system design is generally to decide the local decision rules $\gamma_k(X_k), k=1,\dots, N$, subject to possible constraints imposed on transmission power and bandwidth, so that $P_e$ is minimized. 


When channels are error-free, reporting the raw data or the quantized data to the FC is termed as  {\it centralized detection} (CD)  or {\it distributed detection} (DD), respectively. It is well known that the CD performs better than the DD with a lower Bayes' error, but DD is considered to have an advantage in bandwidth consumption by transmitting less number of bits. The framework of DD was initiated in 1980s by Tenney and Sandell \cite{tenney81:detection} and quickly attracted much attention \cite{tsitsiklis88:decentralized, varshney97:distributed}, although using quantized data for hypothesis testing can be dated back even further \cite{cover69:hypothesis} \cite{kassam77:optimum}. 

When reporting channels are noisy, the FC no longer has the exact observation obtained at sensors, even if sensors report their observed raw data. Therefore, reporting either raw or quantized data both fall under the scope of DD in such a case. One question naturally raised is:  which system performs better? Since CD has better detection performance than DD, one might tend to assume that sending the raw data is still preferred to sending summarized information even if channel noise exists. This was shown incorrect in \cite{chamberland04:asymptotic} in a specific power constrained asymptotic scenario. However, to our best knowledge, no further work has been reported in this area ever since. ``Quantize or Not Quantize" is yet an uncharted area \cite{cao23:distributed}. 

Furthermore, in the scope of conventional DD, codewords are exclusively reported by sensors. In other words,  $Y_k=\gamma_k(X_k)=d, d\in \{0, \dots, D-1\}$ where $D$ is the number of distinct codewords, i.e., short binary bit sequences. The codeword of $d$ indicates the specific partition region in the observation space where the actual observation is located. Any digital modulation method can be used to transmit these codewords. However, an alternative approach emerges when 
sensors transmit specific values, instead of codewords, to the FC, if analog communication is allowed. In communications, by allowing analog modulation, ``{\it To code, or not to code}" has been studied by Gastpar  \cite{gastpar03:code}, and it was shown that the ``uncoded transmission" is appealing as it may lead to an optimal cost-distortion tradeoff with the least possible delay and complexity, and also fits to non-ergodic and multiuser communication scenarios. In \cite{gastpar08:uncoded}, Gastpar further showed that for a simple Gaussian ``sensor" network, uncoded transmission is exactly optimal in the sense that the distortion of the recovered signal, in terms of mean-square error (MSE), decays in $1/N$ while that of the coded system decays in $1/\log_2{N}$, where $N$ is the number of sensors. However, the above work was done using MSE distortion as the performance metric. In the scope of distributed detection when the Bayes' detection error is used as the performance metric,  ``Code or Not Code'', i.e., sending codewords or summarized values, has not been investigated yet. 

In the sequel, for simplicity, we term UDD as the unquantized system that each sensor reports its observation as it is, CDD as the quantized-and-coded system that reports codewords to the FC, and QDD as the quantized-but-uncoded system that reports summarized values to the the FC using analog modulation. While CDD has been the main focus of the past 30$\sim$40 years of history of DD, QDD was only recently proposed in \cite{cao23:distributed}.  Considering channel distortion, existing work in CDD \cite{chen05:optimality,chen09:further,liu06:channel,chau12:coop,cao14:divergence, kasasbeh19:soft, sun22:performance} generally assumes binary channels, with an implicit assumption that codewords can be transmitted via a type of digital modulation, and the channel effect after demodulation can be captured by fixed bit error rates (BERs). This assumption, however, ignores a fact that the transmission power is inherently connected to how the actually quantization is performed, and hence lacks the involvement of power constraint in the design of a CDD system. QDD, on the other hand, can be considered as the extension of CDD, with an additional degree of freedom in the selection of transmission amplitude. As a result, QDD adapts well to the power constraint and hence ``Not Code" is generally preferred to ``Code". Furthermore, QDD makes it possible to answer ``Quantize or Not Quantize?", by comparing with UDD under the same constraints in the transmission power and the bandwidth in term of the number of channel uses. 

The contribution of this paper includes:
\begin{enumerate} 

\item An exposition of QDD and its difference from CDD when addressing the power constraint. 

\item Proofs of a couple of interesting features of QDD. These include the necessary condition for the optimality of using likelihood ratio quantizers (LRQ) in sensor decisions, which is also possessed in CDD. However, in the one-sensor scenario, optimal sensor decision in QDD is in general no longer ``channel blind", which is drastically different from that in CDD. 

\item An attempt to answer "Quantize or Not Quantize ?". Assuming Gaussian noise in both sensing and reporting stages, some comparison results among UDD, CDD and QDD are provided, by considering one-sensor,  two-sensors, and the asymptotically many sensors.

\end{enumerate}

The work to be presented demonstrates that QDD and its comparison with UDD deserves more in-depth work, and could open up a new perspective in distributed detection when both sensing and channel noise exist. 

The rest of the paper is organized as follows. Section II first briefly describes the UDD, CDD and QDD systems.  Then it focuses on CDD with discussion of its sensor decision rules and its insufficiency to consider power constraint. Section III addresses the optimality of using LRQ in sensor decisions in QDD, and the relationship between the LRQ threshold at sensors and the likelihood ratio test (LRT) at the FC. These results can be explored to largely reduce the complexity of searching for the optimal parameters in QDD. Section IV provides some comparison results among UDD, CDD and QDD, by assuming Gaussian noise in both sensing and reporting stages. Finally, Section V concludes the paper and presents some potential directions of future work.

Notation: We use $\Lambda_X(\cdot), \Lambda_Z(\cdot)$ to denote the likelihood ratio at the sensors and at the FC, respectively. 
The subscripts of $X, Z$ could represent a single or multiple random variables, dependent on the context. Let $Z_1^M$ be the short-form of  $\{Z_1,\dots,Z_M\} \in \mathbb{R}^M $, a set of $M$-tuples of real numbers.  Let $Z_{-k}=\{Z_1, \dots, Z_{k-1}, Z_{k+1}, Z_M\} $ be the set of numbers excluding the $k$-th number.  Hence, $Z_1^M = \{Z_k, Z_{-k}\}$. Similarly, the representation applies to $X_1^N$ and $Y_1^N$.  When focusing on a specific sensor in the proofs in Section IV, we generally remove the subscript for simplicity.  

\section{Features and Differences: UDD, CDD and QDD}

\subsection{UDD}

Sensors report observations directly, i.e., $Y_k=X_k, k=1, \dots, N$. Signals received at the FC are $Z_1^M$ via channel $p(Z_1^M|Y_1^N)$. Let $f(z_1^M|H_j)$ be the probability distribution of $z_1^M$ under $H_j, j=0, 1$. The optimal fusion rule for all DD is known as LRT, i.e., 
\begin{eqnarray}
\Lambda_Z(z_1^M)=\frac{f(z_1^M|H_1)}{f(z_1^M|H_0)}\;\mathop{\gtrless}_{H_0}^{H_1}\;  \eta
\label{eqn:udd_lrt}
\end{eqnarray}
where $\eta=\frac{\pi_0}{\pi_1}$. Denote region $R_{Z} =\{ z_1^M\in \mathbb{R}^M |  \Lambda_Z(z_1^M)>\eta \}$. The Bayes' error is 
\begin{eqnarray}
P_{eu} = \pi_1 + \int_{R_{Z}} \left[ \pi_0 f(z_1^M|H_0)-\pi_1 f(z_1^M|H_1) \right] d  z_1^M.
\label{eqn:udd_pe}
\end{eqnarray}

With the assumptions of Gaussian noise and independent reporting channels, we have $M=N$, and the UDD simply becomes testing $\sum_{k=1}^N Z_k$ 
against a threshold $t_z$, and the Bayes' error is
\begin{eqnarray}
p_{eu} = \pi_0 Q\left(\frac{t_z+N  \mu}{\sqrt{N}\sigma}\right) + \pi_1 \left(1-Q\left(\frac{t_z-N  \mu}{\sqrt{N}\sigma}\right) \right)
\label{eqn:centralpe} 
\end{eqnarray} 
where 
\begin{eqnarray}
t_z = \frac{\sigma^2}{2\mu} \ln \eta, \nonumber \\
\sigma^2 = \sigma_s^2+\sigma_c^2, \nonumber \\
\eta=\pi_0/\pi_1, \nonumber \\
 Q(u) = \int_u^{+\infty} \frac{1}{\sqrt{2\pi}} e^{\frac{-v^2}{2}}dv. \nonumber
 \end{eqnarray}

\subsection{CDD, QDD, and Transmission Constraints}

It is known that the decision in FC is always LRT based on the received data. Therefore, the key issue is deciding the sensor decision rules. In CDD, $Y_k=\gamma_k(X_k)=d\in \{0,\dots,D-1\}$, where $Y_k$ is a specific $\log_2 D$-bit-codeword from a set of $D$ distinct codewords. In QDD $Y_k=\gamma_k(X_k)=m_{kd} \in \mathbb{R}, d\in \{0,\dots,D-1\}$. That is, $Y_k$ is a specific real value from a set of $D$ real values specific to sensor $k$. While the set of $D$ codewords in CDD is the same for all sensors, the set of  $\{m_{kd}, d=0, \dots, D-1\}$ in QDD can be completely different for different sensors.

When comparing different systems, the communication cost in bandwidth and power used for each sensor needs to be the same when reporting information to the FC. In CDD, each sensor may report its codeword as a symbol using a specific type of digital modulation. When analog transmission is used as in \cite{gastpar03:code}, the sensed raw data in UDD or the reporting value in QDD can be transmitted as one symbol with a single channel use. Hence, all systems, using either digital or analog modulation, may have the same bandwidth in terms of the number of channel uses.  

When systems use the same number of channel uses per unit of time, equal transmission power means equal average symbol energy per channel use. For UDD, sensor $k, k=1,\dots, N,$ transmits its observation directly. Hence,
\begin{eqnarray}
E_{u,k}= E[X_k^2].
\end{eqnarray}
For QDD, assume that the entire Euclidean space of the observation $X_k$ is partitioned into $D$ intervals with corresponding summarized values as $m_{kj}, j=1, \dots, D$.  The average symbol energy per channel use is 
\begin{eqnarray}
E_{q,k}=\sum_{d=0}^{D-1}p_{m_{kd}}m_{kd}^2 = \sum_{d=0}^{D-1} m_{kd}^2(q_{0kd}\pi_0 +q_{1kd}\pi_1)
\label{eqn:power}
\end{eqnarray}
where  $p_{m_{kd}}$ is the transmission probability of $m_{kd}$. 
$q_{0kd}$ (or $q_{1kd}$) is the probability mass of the $d$-th partition region under $H_0$ (or $H_1$) for the $k$-th sensor. Whenever the observation falls in the $d$-th partition region, $m_{kd}$ is reported. To keep UDD and QDD having the same transmission power, we have $E_{u,k}=E_{q,k}$. 

\subsection{Power issue with CDD}
\label{sec:cdd_power}

For CDD,  a codeword with $\log_2D$ bits can be transmitted as a symbol or multiple symbols with one or multiple channel uses. Existing work of CDD \cite{chen05:optimality,liu06:channel,chau12:coop, cao14:divergence, kasasbeh19:soft, sun22:performance}  exclusively considers BSCs with given BERs. It is implicitly assumed that a digital modulation is used in practice and the relation between the BER and signal-to-noise ratio (SNR) exists for the specific modulation type. This assumption, however, has practical difficulties in detection problems. 

\begin{figure}[htb]
\centering
    \includegraphics[scale=0.30]{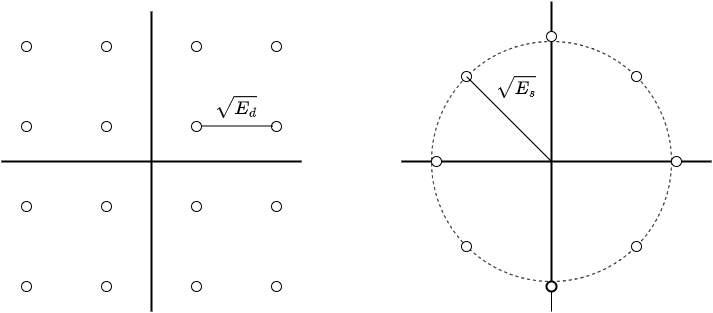}
      \caption{Modulation types: 16-QAM and 8-PSK}
  \label{modulation}
\end{figure}

First, a specific type of modulation, such as M-ary QAM or M-ary PSK, must be used in any practical communication systems. Since each modulation constellation point corresponds to one partition region in the observation space, the probability of using the $i$-th constellation point is determined by the probability mass of the $i$-th region, i.e. $\int_{R_i} f(x|H_0)\pi_0 +f(x|H_1)\pi_1 dx$, where $R_i$ is the region of the observation $X$ where the sensor reports symbol $i$, $ i=1, \dots, \log_2D$. As a result, different constellation points may have different probabilities to be used. In order to keep the same average symbol energy, the minimum distance between two constellation points, such as $\sqrt{E_d}$ in the M-ary QAM constellation shown in Fig. \ref{modulation}, needs to change dynamically as the quantization scheme changes. 

Second, the mapping between the codewords (i.e., the constellation points) and the actual partition regions is critical, even if the constellation does not have to change, such as the M-PSK in Fig. \ref{modulation}, where all constellation points have the same transmission power, irrespective of the quantization schemes. This mapping determines how often one partition region (or symbol) is mistaken as another partition region (or symbol) at the FC due to the channel noise. Different mappings give rise to different transmission probability matrices among the codewords (or the partition regions). This mapping together with how often a specific codeword is used, can make the BERs different in different codewords, an effect that cannot be captured by the BSC assumption with a single BER value. Even for $D=2$, the transition probability from `0' to `1' and that from `1' to `0' may be different, resulting in a non-symmetric binary channel.     

It is worth pointing out even if the BSC assumption with a BER value is valid in any case, the mapping still plays a key role in the final detection performance for $D>2$. In \cite{liu06:channel} the natural binary coding is used directly. In \cite{cao14:divergence}, the codeword mapping was formulated as part of the optimization objective, and the final result disapproves the use of Gray coding as suggested in \cite{chau12:coop}.

Therefore, CDD, that reports codewords only, intends to inform which partition region the observation is located, but has ignored the inherent connection among the constraint of transmission power, the quantization schemes and the codeword mapping. As a result, we have not witnessed yet any comparison between UDD and CDD under the same constraints of transmission power and bandwidth. Nevertheless, the use of codewords in CDD leads to some nice results that are tractable and can largely reduce the design complexity.

\subsection{LRQ of CDD}

One prominent result in CDD is the optimality of using LRQ as the sensor rules when sensor observations are independent. In CDD, $Y_k=\gamma_k(X_k)=d\in \{0,\dots,D-1\}$. In a person-by-person process, denote $C(Z_k, Z_{-k},H)$ as the decision cost at the FC. The local optimal decision at the $k$-th sensor, following \cite{tsitsiklis93:decentralized},  is 
\begin{eqnarray}
\gamma_k^*(X_k) = \arg \min_{\gamma_k(X_k) \in \Gamma_k} E \left [C(Z_k, Z_{-k},H)|X_k \right] \nonumber \\
=\arg \min_{\gamma_k(X_k)\in \Gamma_k} E\left[ E \left[C(Z_k, Z_{-k},H)|H, X_k \right] |X_k\right] \nonumber \\
=\arg \min_{\gamma_k(X_k)\in \Gamma_k} \nonumber \\
E\left[ E\left[ E \left[C(Z_k, Z_{-k},H)|\gamma_k(X_k), H,X_k\right] |H, X_k \right] |X_k\right]
\label{eqn:localdecision}
\end{eqnarray}
where $\Gamma_k$ is set of all possible rules for sensor $k$. 

In \cite{tsitsiklis93:decentralized}, the reporting channel is assumed error-free, i.e., $Z_k = \gamma_k(X_k)$, $k=1,\dots,N$. While in \cite{kashyap06:comments,chen05:optimality}, independent noisy reporting channels are involved through conditional probabilities of $p(Z_k|d), d \in \{0,\dots,D-1\}$. In either case, the key essence of CDD is the use of distinct $D$ codewords and the existence of the Markov property in $X_k\rightarrow d\rightarrow Z_k$. Since $Z_k$ is determined solely by $d$, for simplicity, one can write $E[C(Z_k, Z_{-k}, H)|d, H]$ as $F_k(d, Z_{-k}, H)$. Together with the independence assumption in the reporting channels, for the case of either with or without channel noise, (\ref{eqn:localdecision}) can be simplified as  (\cite{tsitsiklis93:decentralized,kashyap06:comments,chen05:optimality})
\begin{eqnarray}
\gamma_k^*(X_k)= \arg \min_{d=0,\dots, D-1} a_k(H_0,d)P(H_0|X_k) \nonumber \\
+a_k(H_1, d) P(H_1|X_k)
\label{eqn:localDD}
\end{eqnarray}
where $a_k(H_j,d) = E[F_k(d,Z_k,H_j)|H_j]$.  (\ref{eqn:localDD}) shows that the person-by-person optimal sensor decisions are LRQ, but with thresholds determined by $ E[F_k(d, Z_{-k},H_j)|H_j], j=0,1$, and the priori probabilities of hypotheses. This result is addressed in detail in propositions 2.3 and 2.4 in \cite{tsitsiklis93:decentralized} (without channel errors) and proposition 1 in \cite{kashyap06:comments}  and theorem 1 in \cite{chen05:optimality} (with channel errors). The above results assume that the observations at different sensors are independent and the reporting channels between different sensors and the FC are also independent. When each sensor reports one bit decision to the FC, i.e., $D=2$, it was further proved in \cite{chen09:further} that the optimal local sensor decision is still LRQ, even if the channels between sensors and the FC are dependent and noisy, while the sensing observations are independent.

Considering a system with only one sensor and the reporting channel is noisy,  a ``channel blindness'' property was discovered in \cite{liu06:channel} when the Bayes' error is used as the performance metric. This result can be obtained directly from (\ref{eqn:localDD}). We re-state this property and provide a simple proof as follows. 

\textit{\textbf{Lemma 1:}}  For the CDD with only one sensor that sends one bit to the FC where the Bayes' error is used as the performance metric, the optimal decision rule in the sensor is a ``channel-blind'' LRQ. That is, the optimal decision threshold depends on the sensing noise only, and is valid for any channel noise. 

{\it Proof :}  With one sensor,  we have
\begin{eqnarray}
\gamma^*(X) = \nonumber \\
\arg \min_{d=0,1} a(H_0, d) P(H_0|X)+a(H_1,d) P(H_1|X). \nonumber 
\end{eqnarray}
Then, $\gamma^*(X) = 0$ if 
\begin{eqnarray}
   P(H_0|X)(a(H_0,1)-a(H_0,0)) \nonumber \\ 
   \geq P(H_1|X)(a(H_1,0)-a(H_1,1)) 
   \label{eqn:lrtDD}
 \end{eqnarray}
 and  $\gamma^*(X) = 1$ otherwise.  
 
 Here, $a(H_j,d) =E[F(d,H_j)|H_j]$, i.e., the final cost when the sensor decision is $d$ while the true hypothesis is $H_j$. Consider Bayes' error,  
 \begin{eqnarray}
 a( H_j,d) = \int_{Z}I(\gamma_0(z)\neq H_j)p(z|d)dz 
 \end{eqnarray}
 where $I(\cdot)$ is the indication function and $\gamma_0(\cdot)$ is the FC decision rule. Denote $R_Z = \{z|\gamma_0(z) = H_1\}$ as the region of $Z$ that decides $H_1$. We have
  \begin{eqnarray}
 a( H_0,d) = \int_{R_Z}p(z|d)dz =1-a(H_1,d).
 \end{eqnarray}
 
 Plugging back to (\ref{eqn:lrtDD}), $\gamma^*(X) = 0$ when 
 \begin{eqnarray}
   P(H_0|X)(a(H_1,0)-a(H_1,1)) \nonumber \\ 
   \geq P(H_1|X)(a(H_1,0)-a(H_1,1)) .
 \end{eqnarray}
 Since $a(H_1,0)>a(H_1,1)$ due to the definition of cost, 
 the local decision is simply the LRT test of $p(X|H_1)/p(X|H_0)$ against $\eta$, irrespective of the channel $p(Z|d)$, and hence the LRQ is ``channel blind''. 
 {\flushright $\square$}

These features apparently can be exploited to reduce the complexity in searching the suboptimal CDD systems.  
 
\section{Results of QDD}
 
\subsection{LRQ of QDD}

The fundamental feature of QDD is that we allow the transmission of real values instead of codewords. That is, $Y_k = \gamma_k(X_k)=m_{kd}, d=0, \dots, D-1$. Let $\Gamma_k^*$ be the set of all possible $\gamma_k(X_k)$ rules. Determining $\gamma_k(X_k)$ includes two processes. One is to partition the observation space $\mathbb{R}$, where the observation $X_k$ belongs to, into $D$ regions. The other is to determine the $D$ values of $m_{kd}, d=0, \dots, D-1$ that are to be transmitted when the observation falls in each of the $D$ different regions. The value of $m_{kd}$ is reported to the FC via analog modulation and all the $D$ values must satisfy the power constraint of (\ref{eqn:power}).  In CDD,  the same $D$ codewords are always used, no matter how the observation space is partitioned. In QDD, on the contrary, $\{m_{kd}, d= 1, \dots, D\}$ could be any set of values as long as they meet the power constraint. Given specific partitions, i.e., a quantization scheme, $\{p_{m_{kd}}, d=0, \dots, D-1\}$ are determined, but the selection of $m_{kd}$ satisfying (\ref{eqn:power}) is not unique. Moreover, changing the partitions to any degree also affects $p_{m_{kd}}$ and hence the selection of $\{m_{kd}, d=0, \dots, D-1\}$. This further changes the distribution of $Z_1^M$ at the FC. 

As a result, finding $\gamma_k(X_k)$ is a complex optimization process. It is beneficial if the search space of $\Gamma_k^*$ could be largely reduced. In the following, we prove that with a person-by-person process, for QDD, LRQ is still a necessary condition for the optimality of sensor decision rules when $D=2$, regardless of whether the reporting channels are noisy or not, and independent or not.   

 \textit{\textbf{Theorem 1:}}  For a QDD system where two possible values are used to report observations in each sensor, when $X_k, k=0, \dots, N-1$ are independent, a necessary condition for the optimal decision rule of each sensor is the LRQ. Furthermore, this result is irrelevant to the assumption of reporting channels. That is, the channel function could be in any form of $p(Z_1^M|Y_1^N)$. This result is also irrelevant to which decision rule is used at the FC. The optimality holds under both Neyman-Pearson (NP) and Bayesian criteria.  
   
{\it Proof :} 
Consider a specific sensor while decision rules of other sensors and the fusion rule are fixed, our approach is to show that for the sensor, the LRQ rule could perform at least as good as any non-LRQ rule, with less or equal power that is used by the non-LRQ rule. Consider two possible decision rules to be used in the sensor: One is a non-LRQ rule and the sensor reports either $m_0$ or $ m_1$ to the FC, where $m_0, m_1$ satisfy the power constraint (\ref{eqn:power}). The other is a LRQ rule. Suppose that this LRQ rule also uses the same $m_0$ and $m_1$ as the non-LRQ rule used for reporting. Denote $P_f=P(m_1|H_0), P_d=P(m_1|H_1)$ as the false alarm and detection probabilities at the sensor for the non-LRQ rule, and $P_f^*, P_d^*$ as the false alarm and detection probabilities for the LRQ rule. Based on the Neyman-Pearson's Lemma, for any non-LRQ rule, we can always find a LRQ rule such that $P_d^*\geq P_d$ if $P_f^*=P_f$, or $P_f^*\leq P_f$ if $P_d^*=P_d$. 

Under the condition that hypothesis $H_0$ is true, the average symbol energy for the two systems are as follows. 
\begin{eqnarray}
E_T = (1-P_f)m_0^2+P_f m_1^2, \nonumber \\
 E_T^* = (1-P_f^*)m_0^2 + P_f^*m_1^2. 
\end{eqnarray}
Under the condition that hypothesis $H_1$ is true, we have  
\begin{eqnarray}
E_T = (1-P_d)m_0^2+P_d m_1^2, \nonumber \\
E_T^* = (1-P_d^*)m_0^2 + P_d^*m_1^2.
\end{eqnarray}

If $m_0^2 \leq m_1^2$, suppose that $P_d^*=P_d$, we have $E_T=E_T^*$ under $H_1$ but $E_T^*\leq E_T$ under $H_0$. If $m_0^2>m_1^2$, suppose $P_f^*=P_f$, we have $E_T=E_T^*$ under $H_0$ but $E_T^*\leq E_T$ under $H_1$.  In other words, for any non-LRQ rule, we can always find a LRQ rule that uses the same $m_0, m_1$ but with less or equal transmission power compared to the non-LRQ rule. It must be noted that $m_0, m_1$ sent in the LRQ rule might not be the optimal values that lead to the best performance. 

Next, we follow the approach used in \cite{chen09:further} to show the optimality of LRQ at sensors under both NP and Bayesian criteria at the FC. Let $U_0=\gamma_0(Z_1^M)$ be the final decision at the FC, and $U_0=j$ decides $H_j, j=0, 1$. Let $P_F$ and $P_D$ are the false alarm and detection probabilities after the FC has made a decision. When the non-LRQ rule is used in sensor $k$, then
\begin{eqnarray}
P_F = P(U_0=1|H_0) = P(U_0=1, Y_k=m_0|H_0) \nonumber \\
+ P(U_0=1, Y_k=m_1|H_0) \nonumber \\
=P(U_0=1|Y_k=m_0,H_0) + a_k P(Y_k=m_1|H_0)\nonumber \\
=P(U_0=1|Y_k=m_0,H_0) + a_k P_f
\label{eqn:pf}
\end{eqnarray}
where 
\begin{eqnarray}
a_k =\nonumber \\
 P(U_0=1|Y_k = m_1, H_0)-P(U_0=1|Y_k=m_0,H_0).
\end{eqnarray}

Similarly,
\begin{eqnarray}
P_D = P(U_0=1|H_1) = P(U_0=1, Y_k=m_0|H_1) \nonumber \\
+ P(U_0=1, Y_k=m_1|H_1) \nonumber \\
=P(U_0=1|Y_k=m_0,H_1) + b_k P(Y_k=m_1|H_1)\nonumber \\
=P(U_0=1|Y_k=m_0,H_1) + b_k P_d
\label{eqn:pd}
\end{eqnarray}
where 
\begin{eqnarray}
b_k = \nonumber \\
P(U_0=1|Y_k = m_1, H_1)-P(U_0=1|Y_k=m_0,H_1).
\end{eqnarray}

In the above, only $P_f$ and $P_d$ are related to the decision rule used in sensor $k$. Other terms, including $a_k,b_k, P(U_0=1|Y_k=m_0, H_0), P(U_0=1|Y_k=m_0, H_1)$, are only related to the feature of the channel and the fusion rule. It is left to show that we can always find a LRQ rule for sensor $k$ that makes the final performance at the FC, i.e.,  $(P_F^*, P_D^*)$, improved over that of the non-LRQ rule. 
\begin{enumerate}
\item When $a_k>0, b_k>0$, a LRQ rule can be found that $P_f^* \leq P_f$ and $P_d^* \geq P_d$. Putting back to (\ref{eqn:pf})(\ref{eqn:pd}), we have $P_F^*\leq P_F$ and $P_D^*\geq P_D$.

\item When $a_k>0, b_k<0$, we may pick a LRQ rule with threshold to be $\infty$, i.e., the sensor always send $m_0$. Then $P_d^*=0$ and $P_f^*=0$. Putting back to (\ref{eqn:pf})(\ref{eqn:pd}), it gives a performance of $(P_F^*, P_D^*)$ better or equal to $(P_F, P_D)$. 

\item When $a_k<0, b_k>0$, we may pick a LRQ rule with threshold to be $0$, i.e., the sensor always send $m_1$. Then $P_d^*=1$ and $P_f^*=1$. Putting back to (\ref{eqn:pf})(\ref{eqn:pd}), it gives a performance of $(P_F^*, P_D^*)$ better or equal to $(P_F, P_D)$. 

\item When $a_k<0, b_k<0$, we may use the LRQ in 1) but switch the decision of $U_0$ from `0' to `1' and from `1' to `0', which is equivalent to minimizing (maximizing) the sensor detection probability (false alarm probability).  Putting back to (\ref{eqn:pf})(\ref{eqn:pd}), it gives a performance of $(P_F^*, P_D^*)$ better or equal to $(P_F, P_D)$. 

\end{enumerate} 

As a result, consider sensor $k$ while other sensor decisions and fusion rules are fixed, for any non-LRQ decision rule, we can always find a LRQ rule that uses the same $m_0, m_1$ as the non-LRQ rule uses, but gives the same or better detection performance with at most the same transmission power used by the non-LRQ rule. In addition, a LRQ rule with optimized $m_0, m_1$ under the power constraint cannot perform worse than the one simply adopting the $m_0, m_1$ values used by the non-LRQ rule. Therefore, the optimality of LRQ sensor rules has been established under the Neyman-Pearson's criterion. 

Under the Bayesian criterion, the detection error probability  $P_e = \pi_0 P_F +\pi_1(1-P_D)$. Using (\ref{eqn:pf})(\ref{eqn:pd}), we have
\begin{eqnarray}
P_e = c_k +a_k\pi_0 P_f - b_k \pi_1 P_d
\end{eqnarray}
where $c_k = \pi_0P(U_0=1|Y_k = m_0, H_0) +\pi_1(1-P(U_0=1|Y_k = m_0, H_1))$ is a value irrelevant to the sensor decision rule.  
As a result, following the four cases of $a_k$ and $b_k$ and the proof for the NP criterion, it shows directly for any non-LRQ rule, we can always find a LRQ rule with the detection error probability $P_e^*\leq P_e$. 
  
{\flushright $\square$}
 
 This result reduces the search space of $\Gamma_k^*$ to the set of LRQ rules. 
    
\subsection{One-Sensor Case - Sensor Rule vs. Fusion Rule}
 
We consider a single sensor scenario where  $Y= \gamma(X)=m_1$ when $\Lambda_X(x)\geq \lambda$ and $Y_k=m_0$ otherwise. $\lambda$ is a threshold in the LRT using $X$. The value of $\lambda$ determines a pair of false alarm and detection probabilities, i.e., $P_f $ and $P_d$, at the sensor.  At the FC, $Z = Y+ W$, where $W$ is the additive noise following a distribution $f_W(w)$. Let $f(Z|H_j)$ be the distribution of $Z$ conditioned on $H_j, j=0,1$. LRT is then applied to $Z$ to make a final decision at the FC, i.e., 
\begin{eqnarray}
\Lambda_{Z}(Z)=\frac{f(Z|H_1)}{f(Z|H_0)} \;\mathop{\gtrless}_{H_0}^{H_1} \eta.
\label{eqn:lrt_1s}
\end{eqnarray}
 
The optimal parameters of the sensor decision rule need to be determined to minimize the detection error at the FC. We establish a result that links the LRT threshold at the FC and the LRQ sensor rule with $(P_f, P_d)$, $m_0$ and $m_1$ as follows. 
 
 \textit{\textbf{Lemma 2:}}  Consider the one-sensor case where the sensor uses a LRQ rule that transmits $Y=m_i$ whenever the observation falls in the $i$-th region, $i=0, 1$. Let $P_f=P(m_1|H_0)$ and $P_d=P(m_1|H_1)$, where $P_d \geq P_f$. Let the received signal at the FC be $Z=Y+W$, where $W$ is the additive noise with a distribution of $f_W(w)$. Let $\Lambda_W(z) = \frac{f_W(z-m_1)}{f_W(z-m_0)}$ and the final detection error at the FC be $P_e$. Then, the optimal decision at the FC is that 1) when $\eta \geq \frac{P_d}{P_f}$, it always decides $H_0$ and $P_e=\pi_1$; 2) when $\eta \leq \frac{1-P_f}{1-P_d}$, it always decides $H_1$ and $P_e=\pi_0$;  3) when $\frac{1-P_f}{1-P_d} < \eta <\frac{P_d}{P_f}$, it decides based on 
 \begin{eqnarray}
 \Lambda_W(z) \;\mathop{\gtrless}_{H_0}^{H_1} 1+\frac{\eta-1}{P_d-\eta P_f} \triangleq \lambda_{\eta}, \nonumber
 \end{eqnarray}
 and $P_e = \pi_1+ \int_{\lambda_{\eta}}^\infty \left[ \pi_0 f_{\Lambda_W|H_0}(\lambda) - \pi_1 f_{\Lambda_W|H_1}(\lambda) \right]  d\lambda$, where $f_{\Lambda_W|H_j}(\lambda)$ is the probability distribution of $\Lambda_W$ under $H_j$, $j=0,1$.
 
{\it Proof :}  Given $Y=m_i$, the distribution of the received signal at the FC is $f_{Z|m_i}(z) = f_W(z-m_i), i=0, 1$. LRT at the FC, i.e., (\ref{eqn:lrt_1s}), becomes 
\begin{eqnarray}
 \frac{P_d f_W(z-m_1)+(1-P_d)f_W(z-m_0)}{P_f f_W(z-m_1)+(1-P_f) f_W(z-m_0)} \;\mathop{\gtrless}_{H_0}^{H_1}  \eta.
 \end{eqnarray}
 Equivalently, 
 \begin{eqnarray}
 \frac{P_d \Lambda_W(z)+(1-P_d)}{P_f \Lambda_W(z)+(1-P_f))} \;\mathop{\gtrless}_{H_0}^{H_1}  \eta
 \end{eqnarray}
 which further gives
 \begin{eqnarray}
 (P_d -\eta P_f) \Lambda_W(z) \;\mathop{\gtrless}_{H_0}^{H_1}  (P_d-\eta P_f)+ (\eta-1). 
 \label{eqn:gamma}
 \end{eqnarray} 
 
 When $\eta > \frac{P_d}{P_f}$, we have $P_d-\eta P_f <0$. (\ref{eqn:gamma}) becomes
 \begin{eqnarray}
  \Lambda_W(z) \;\mathop{\gtrless}_{H_1}^{H_0}  1+\frac{\eta-1}{P_d-\eta P_f} \nonumber
  \label{eqn:case2}
 \end{eqnarray}  
 where  the right-hand side is negative when $\eta>\frac{1-P_d}{1-P_f}$, which is always true because $\eta>\frac{P_d}{P_f}$. As a result, the optimal decision at the FC is always deciding $H_0$. Consequently, $P_e = \pi_1$. When $\eta=\frac{P_d}{P_f}$, the left-hand side of (\ref{eqn:gamma}) is zero while the right-hand side is a non-negative value considering $\eta=P_d/P_f \geq 1$. Again, FC decides $H_0$ and $P_e = \pi_1$. 
  
 When $\eta<\frac{P_d}{P_f}$, we have $P_d-\eta P_f >0$. Then, 
 \begin{eqnarray}
  \Lambda_W(z) \;\mathop{\gtrless}_{H_0}^{H_1}  1+\frac{\eta-1}{P_d-\eta P_f} 
  \label{eqn:case3}
 \end{eqnarray}  
where the right-hand side is a non-positive value when $\eta \leq \frac{1-P_d}{1-P_f}$. As a result, FC always decides $H_1$ when $\eta \leq \frac{1-P_d}{1-P_f}$ and consequently $P_e= \pi_0$. 

When $ \frac{1-P_d}{1-P_f} < \eta <\frac{P_d}{P_f}$, $\lambda_{\eta}$ is a positive value and FC decides based on (\ref{eqn:case3}). Consequently, 
\begin{eqnarray}
P_e = \pi_1+ \int_{\lambda_{\eta}}^\infty \left[ \pi_0 f_{\Lambda_W|H_0}(\lambda) - \pi_1 f_{\Lambda_W|H_1}(\lambda) \right]  d\lambda. 
\end{eqnarray}

{\flushright $\square$}

It should be noted that {\it Lemma 2} is valid for any type of additive noise distribution $f_W(w)$. Given this distribution and the parameters $P_d, P_f, \pi_1, m_i, i=0,1$, the threshold of LRT at the FC can be determined immediately. 

{\it Example of Gaussian noise case of $(\sigma_s, \sigma_c)$}: Assume that sensing and reporting both experience AWGN with standard deviation of $\sigma_s$ and $\sigma_c$, respectively. In such a case, it is convenient to consider the thresholds $t, t_z$ in the domains of $X$ and $Z$ directly, instead of the thresholds of the likelihood ratios. That is, $Y=m_1$ when $X\geq t$ and $Y=m_0$ when $X<t$. At the FC, decision is made as  
\begin{eqnarray}
 z \mathop{\gtrless}_{H_0}^{H_1} t_z.\nonumber
 \end{eqnarray}

Given the sensor decision parameters, we have $P_d = Q\left(\frac{t-\mu}{\sigma_s} \right)$, and $P_f = Q\left(\frac{t+\mu}{\sigma_s} \right) $. From {\it Lemma 2}, when $\eta \geq \frac{P_d}{P_f}$, $t_z = \infty$ and $P_e=\pi_1$. When $\eta \leq \frac{1-P_d}{1-P_f} $, $t_z = -\infty$ and $P_e=\pi_0$. When $\frac{1-P_d}{1-P_f} < \eta <\frac{P_d}{P_f}$,  
 \begin{eqnarray}
 t_z =  \frac{\sigma_c^2}{m_1-m_0}\ln \lambda_{\eta}+\frac{m_0+m_1}{2}.
 \label{eqn:t_z}
 \end{eqnarray}
 Hence, 
 \begin{eqnarray}
 P_e = \pi_1+\pi_0 P_F - \pi_1 P_D 
\label{eqn:1s_pe}
 \end{eqnarray} 
 where
  \begin{eqnarray}
  \label{eqn:PF}
 P_F = Q\left(\frac{t_z-m_0}{\sigma_c} \right)(1-P_f) + Q\left(\frac{t_z-m_1}{\sigma_c} \right)P_f , \\
 \label{eqn:PD}
 P_D = Q\left(\frac{t_z-m_0}{\sigma_c} \right)(1-P_d) + Q\left(\frac{t_z-m_1}{\sigma_c} \right)P_d. 
 \end{eqnarray}  
  
Moreover, $m_0, m_1, t$ are linked by the power constraint (\ref{eqn:power}). That $E_{q} = E_u$ gives $m_0^2 p_{m_0} + m_1^2 p_{m_1} =\mu^2+\sigma_s^2 $ where $p_{m_1} = P_d \pi_1 +P_f\pi_0$ and  $p_{m_0} =1-P_{m1}$. 
 
Therefore, in order to minimize the final $P_e$, the actual unknown variables are only $m_0$ and $t$ which can be found via many popular optimization solvers. A simple enumeration process could also quickly provide a solution with high accuracy.  It must be pointed out that the optimal $t$ connects to $m_0$ and $m_1$, which further connect to $P_e$. In general, $t$ is no longer ``channel blind", and its value must be solved via an optimization process, which is drastically different from CDD. 

A special case, however, is when $\eta=1$, i.e., $\pi_0=\pi_1=0.5$. We can find $\lambda_\eta=1$, and from (\ref{eqn:t_z}), $t_z = \frac{m_0+m_1}{2}$. Putting this back to the calculation of $P_F$ and $P_D$ in (\ref{eqn:PF})(\ref{eqn:PD}), then (\ref{eqn:1s_pe}) becomes 
\begin{eqnarray}
P_e = 0.5 + 0.5 (P_d - P_f)\left[ 2Q\left(\frac{m_1-m_0}{2\sigma_c} \right)-1 \right]. 
\end{eqnarray}
For any $m_0,m_1$,  from $dP_e/dP_f=0$, we see immediately that $dP_d/dP_f = 1$, which means that the local sensor decision is the LRQ at $\eta = 1$. Hence, we have $t=0$. Consequently, we have $p_{m_1}=p_{m_0}=0.5$ and $m_0^2+m_1^2 = 2E_u$. Now from $dP_e/dm_0=0$, we have $dQ\left(\frac{m_1-m_0}{2\sigma_c}\right) /dm_0 = 0$. By using the power constraint, it further gives that $m_0^2=E_u$. As a result, we can conclude that, for the 1-sensor case with equal priors, the optimal local sensor decision rule is that $t=0, m_0=-m_1=-\sqrt{E_u}$, which is ``channel blind" in this special case. Consequently, from (\ref{eqn:1s_pe}), the Bayes' error can be found as
\begin{eqnarray}
P_e = Q \left(\frac{\sqrt{E_u}}{\sigma_c}\right) + Q \left(\frac{\sqrt{\mu}}{\sigma_s}\right) - 2Q \left(\frac{\sqrt{E_u}}{\sigma_c}\right)Q \left(\frac{\sqrt{\mu}}{\sigma_s}\right).
\label{eqn:pe_1s}
\end{eqnarray}

 \subsection{Optimization for QDD Design}
 
Consider a QDD system with $N$ sensors that have independent observations via AWGN channel, i.e., $X_k \sim \mathcal{N}(-\mu, \sigma_{s_k})$ under $H_0$ and $X_k \sim \mathcal{N}(\mu, \sigma_{s_k})$ under $H_1$, $k=1, \dots, N$. For each sensor, $D=2$ is used.  Let $P_{eq}$ be the final decision error when the FC uses LRT. $P_{eq}$ and the LRT still follow (\ref{eqn:udd_lrt}) and (\ref{eqn:udd_pe}) except that $z_1^M$ are the received signals in QDD instead of UDD.  Based on {\it Theorem 1}, the design of the QDD system becomes an optimization process that considers only the set of LRQ sensor decision rules as follows. 
\begin{equation}
\begin{aligned}
P_1: \quad \min_{\substack{t_k, m_{k0},m_{k1}, \\ k=1,\dots,N}} &  \quad  P_{eq} \\
\textrm{s.t.} \quad & -\sqrt{\frac{E_{u,k}}{p_{m_{k0}}}}\leq m_{k0} \leq \sqrt{\frac{E_{u,k}}{p_{m_{k0}}}}\\
	  &p_{m_{k0}}m_{k0}^2 + p_{m_{k1}}m_{k1}^2 = E_{u,k} \nonumber \\
	  &p_{m_{k1}} =\pi_0 Q\left(\frac{t_k+\mu}{\sigma_{s_k}}\right) + \pi_1Q\left(\frac{t_k-\mu}{\sigma_{s_k}}\right) \\
	  &p_{m_{k0}}=1-p_{m_{k1}}	  
\end{aligned}
\end{equation}

Sensing observations should be independent but need not to be identically distributed. Numerical evaluation of the final $P_{eq}$ based on the LRT fusion rule is needed, which must consider the channel model. The LRT needs to compare $\Lambda_Z(z_1^N)$ against $\eta$ over all possible $z_1^N$ values. When reporting channels are independent, {\it Lemma 2} may be utilized to reduce the computational complexity. For example, when $N=2$, given the value of $z_1$,  deciding $\Lambda_{Z}(z_1, z_2) \;\mathop{\gtrless}_{H_0}^{H_1} \eta$ is equivalent to deciding $\frac{f_2(z_2|H_1)}{f_2(z_2|H_0)} \;\mathop{\gtrless}_{H_0}^{H_1} \eta^*$, where $\eta^* = \eta \frac{f_1(z_1|H_0)}{f_1(z_1|H_1)} $, for which {\it Lemma 2} can be used directly to find the region of $z_2$ that the FC decides $H_1$, given the value of $z_1$. This eliminates the need to calculate $\Lambda_Z(z_1, z_2)$ for each individual $(z_1,z_2)$. Here, $f_k(z_k|H_j)$ is the probability distribution of $Z_k$ received from the $k$-th sensor under $H_j$, where $k=1, 2$ and $j=0,1$.

When $N$ is large, solving the optimization problem becomes complicated. It needs numerical evaluation of $f(z_1, \dots, z_M|H_j)$, $j = 0,1$, for all possible $z_1^M$, for each possible group of values of $t_k, m_{k0}, m_{k1}, k=1,\dots,N$, in the optimization process. The complexity could be very high even for a moderate $N$ value. However, when i.i.d. noise is assumed in both the sensing stage and the reporting stage, the process can be sped up by assuming that all $N$ sensors are using the same group of $\{t, m_{0}, m_{1}\}$ values. The above optimization problem is formulated with the assumption of using AWGN noise. It can also be re-formulated when other types of noise are considered.

\section{Comparison and Numerical Results}

Assuming AWGN for both sensing and reporting stages and considering $D=2, \mu=1$, we compare the performance of UDD, CDD and QDD. For QDD, optimal sensor decision rules are obtained via solving the optimization problem $P_1$. For CDD, sensor $k$ always transmits either $\sqrt{E_{u,k}}$ or $-\sqrt{E_{u,k}}$ using BPSK, but with the optimal LRQ value. This is equivalent to solving the optimization problem $P_1$ by fixing $m_{k0}=-\sqrt{E_{u,k}}$ and $m_{k1}=\sqrt{E_{u,k}}$, for $k=1,\dots,N$.  At the FC, LRT rule is used to calculate the Bayes' errors of different systems for comparison. 

\subsection{One-sensor case}
 
\begin{figure}[htbp]
	\centerline{\includegraphics[width=.44\textwidth]{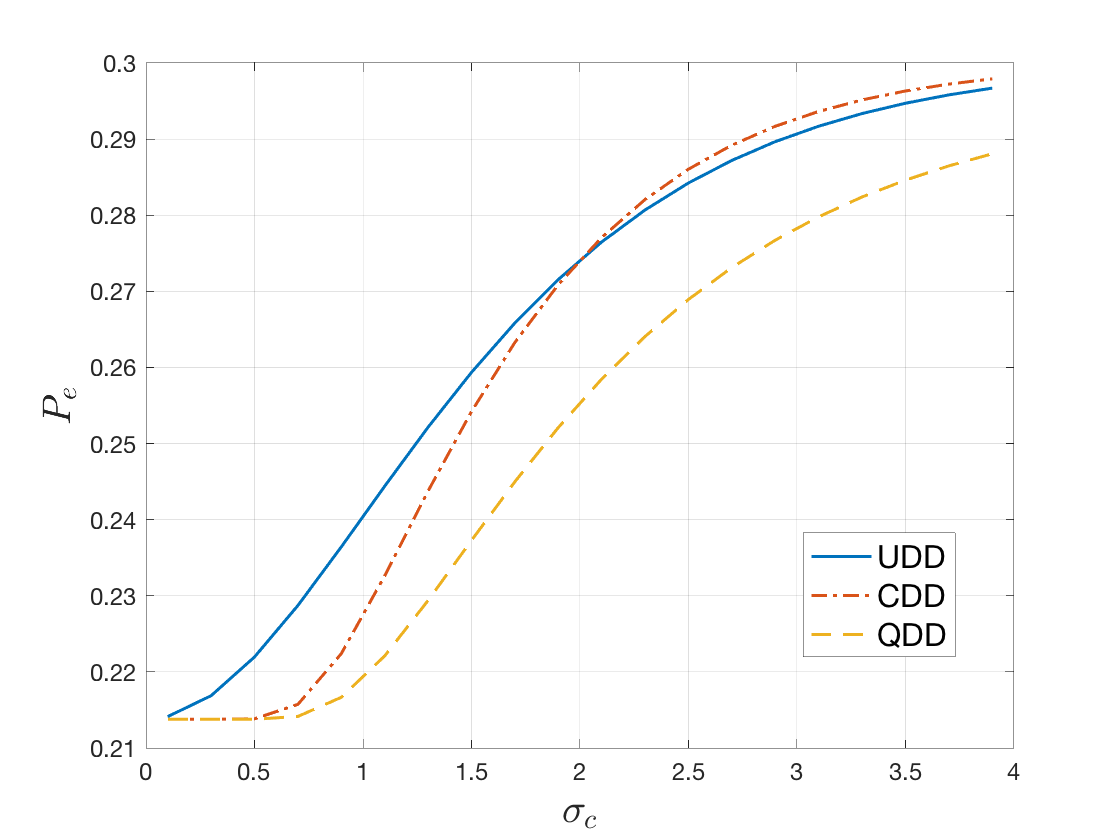}}
	\caption{$\pi_1=0.7, \mu=1, \sigma_s=1.5$, UDD, CDD and QDD, 1-sensor and $D=2$.}
	\label{fig:p1_07_pe}
\end{figure}

Fig. \ref{fig:p1_07_pe} shows the Bayes' error $P_e$ for the UDD, CDD and QDD, when $\pi_1=0.7$, $\sigma_s =1.5$ while $\sigma_c$ changes. It can be observed that QDD performs better than CDD as it has an additional degree of freedom in parameter selection. In the test scenario, compared with UDD, CDD performs better when $\sigma_c<2.1$ but worse when $\sigma_c$ increases. QDD shows consistent improvement over both UDD and CDD. Performance of the three systems all converge to the same value as the channel noise goes to zero. 

\begin{figure}[htb]
	\centerline{\includegraphics[width=.44\textwidth]{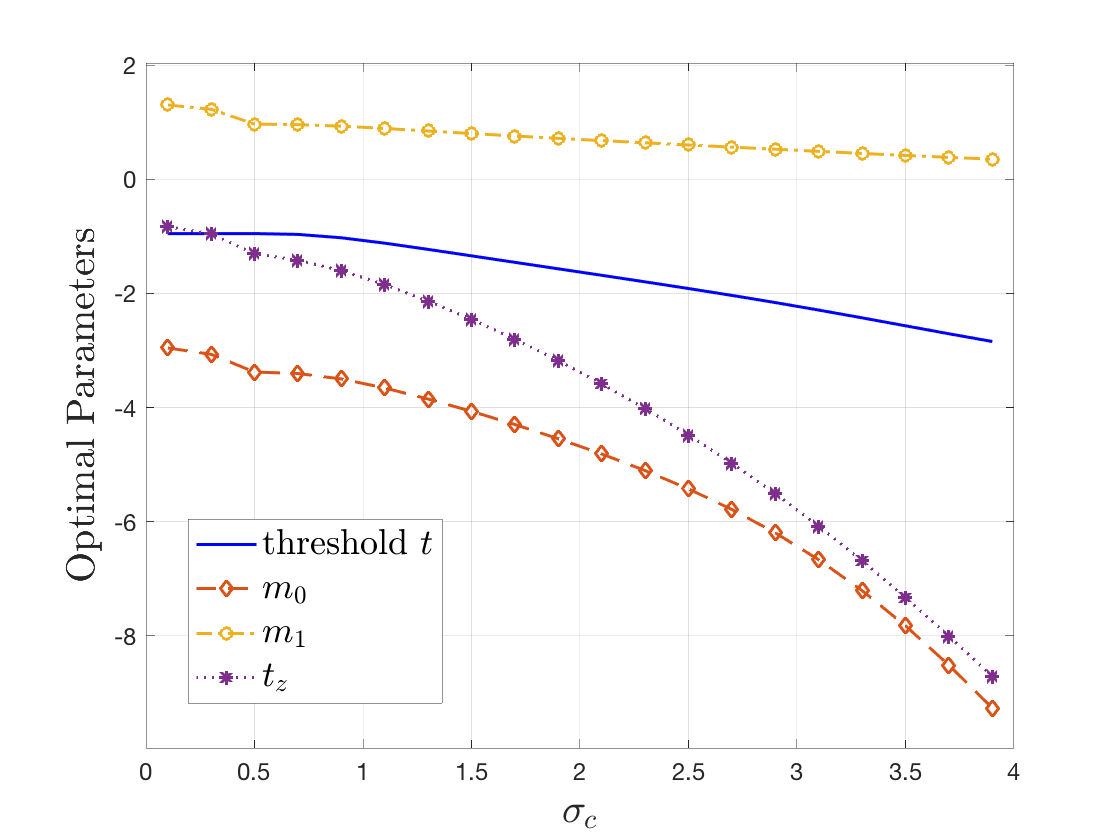}}
	\caption{$\pi_1=0.7, \mu=1, \sigma_s=1.5$, sensor rules of QDD, 1-sensor and $D=2$. }
	\label{fig:p1_07_para}
\end{figure}

Fig. \ref{fig:p1_07_para} shows the optimal sensor decision values of $t, m_0, m_1$ in QDD obtained by solving the optimization problem $P_1$ for the above case. For QDD, the detection threshold value $t_z$ at the FC for the finally received signal $Z$ is also plotted. The sensor decision threshold $t$ in QDD clearly depends not only on $\sigma_s$ but also on $\sigma_c$, which shows that the best sensor decision that minimizes the local sensing error does not give the optimal system performance when the channel distortion is present.

For the CDD case, the threshold in the sensor is channel blind and was set as $\frac{\sigma_s^2}{2\mu} \log \frac{\pi_0}{\pi_1}$ always. Bits of `1' and `0' were transmitted using BPSK with bit energy of $\sqrt{E_u}$, i.e., $m_1 =-m_0 = \sqrt{3.25}$. It is worth noting that when $D>2$, it is difficult to compare CDD and UDD due to the dynamic relationship among the quantization, the mapping between the constellation points and the partition regions, and the power constraint as discussed in Section \ref{sec:cdd_power}.

\vspace{-0.08in}

\subsection{Two-Sensors Case with i.i.d. Noise}

\begin{figure}[htbp]
	\centerline{\includegraphics[width=.44\textwidth]{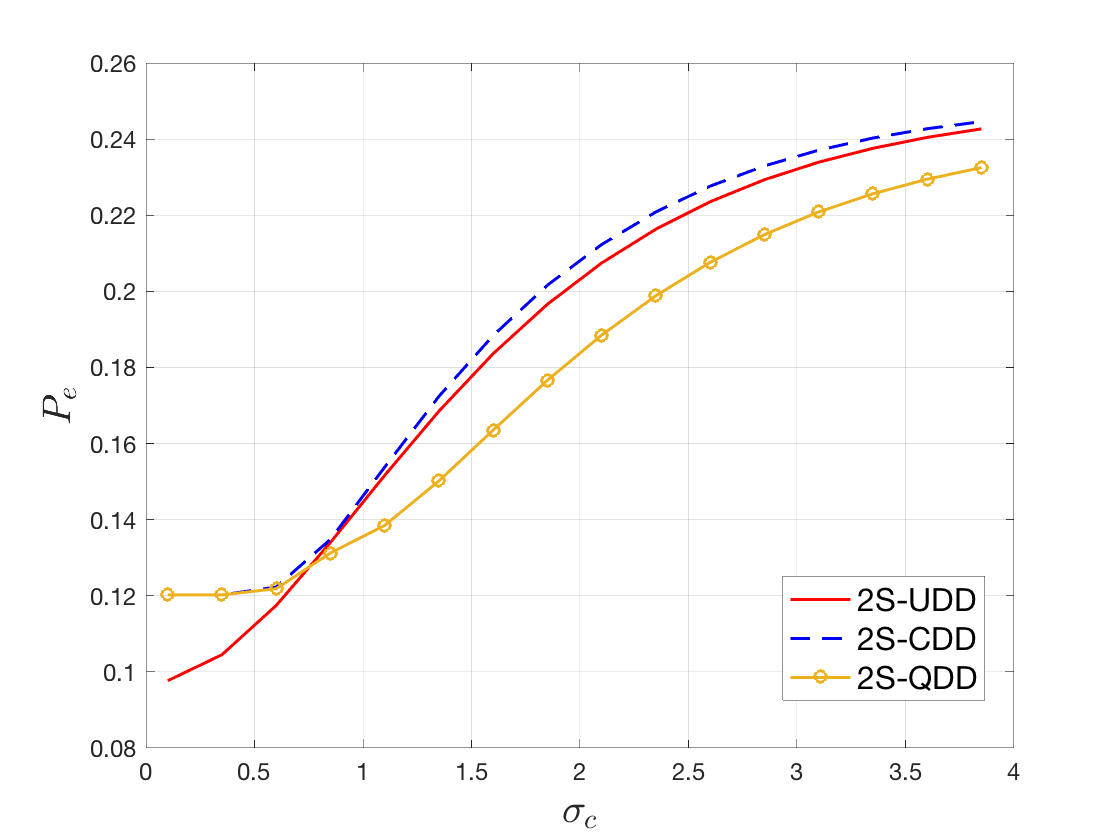}}
	\caption{Bayes error $P_e$. $\pi_1=0.75, \mu=1, \sigma_s=1.2$, 2-sensors, $D=2$ }
	\label{fig:p1_75_2s_pe}
\end{figure}

Consider a scenario where $\pi_1 = 0.75$, $\sigma_s=1.2$ and $\sigma_c$ changes. Both sensors assume i.i.d. AWGN in the sensing stage and i.i.d. AWGN in the reporting stage.  Fig. \ref{fig:p1_75_2s_pe} shows the comparison of the detection error for the 2-sensors case. `2S-CDD' case uses BPSK with $m_{k1}=-m_{k0}=\sqrt{E_u}$, $k=1,2$. However, the optimal sensor thresholds are searched optimally. For the `2S-QDD' case, it searches the optimal values for all $t_k, m_{k0}, m_{k1}, k=1,2$, that minimize the decision error. From Fig. \ref{fig:p1_75_2s_pe}, it can be observed that CDD always performs worse than UDD in this scenario. While QDD performs worse than UDD but performs very similarly to CDD when $\sigma_c<0.85$, it consistently performs better than UDD and CDD as $\sigma_c$ increases. It is worth noting that when $\sigma_c$ approaches to zero, unlike the one-sensor case, UDD always performs better than QDD and CDD. This is because in such a case, UDD approaches to CD, which is always better than DD in the performance of detection when multiple sensors are used. 

\begin{figure}[htbp]
	\centerline{\includegraphics[width=.44\textwidth]{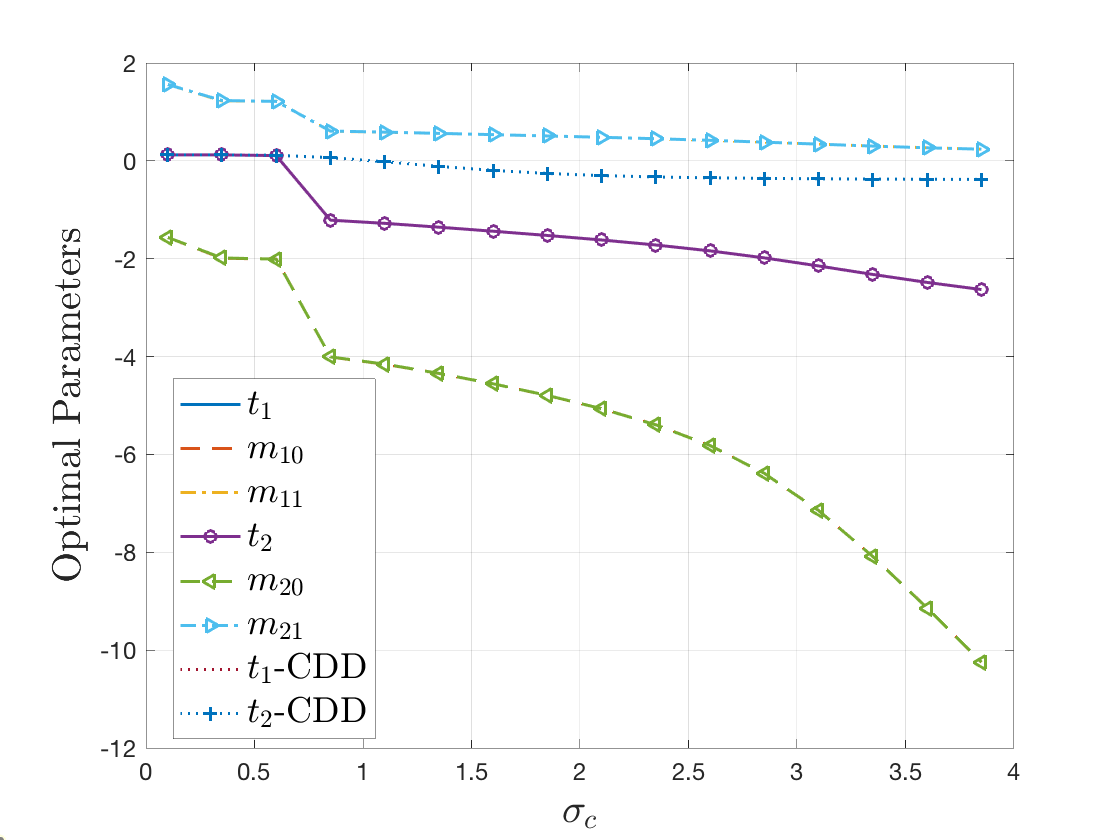}}
	\caption{Optimal sensor rules. $\pi_1=0.75, \mu=1, \sigma_s=1.2$, 2-sensors, $D=2$. }
	\label{fig:p1_75_2s_para}
\end{figure}

Fig. \ref{fig:p1_75_2s_para} shows the optimal parameters for both CDD and QDD. In this scenario, the obtained optimal sensor decision rules are the same for the two sensors, both in CDD and in QDD. It can also be observed that with two sensors, the sensors' LRQ thresholds in both CDD and QDD are not channel blind.  

 
\subsection{Two-Sensors Case with Correlated Channel Noise}
\label{correlated}

{\it Theorem 1} only requires independence in the sensor observations which need not to be identically distributed. Furthermore, it has no restriction on the model of reporting channels. In the following, we assume the channel model is a bivariate Gaussian noise with zero means, i.e., $(W_1, W_2) \sim G(0, 0, \rho, \sigma_{c_1}, \sigma_{c_2})$. 

At the FC, for UDD we have $Z_k = X_k+W_k, k=1,2$. Therefore, $(Z_1, Z_2) \sim G(s, s, \hat{\rho}, \sigma_1, \sigma_2)$ under $H_1$, and  $(Z_1, Z_2) \sim G(-s, -s, \hat{\rho}, \sigma_1, \sigma_2)$ under $H_0$, 
where $\sigma_k = \sqrt{\sigma_{s_k}^2+\sigma_{c_k}^2}$, $k=1, 2$, and $\hat{\rho} = \frac{\rho \sigma_{c_1}\sigma_{c_2}}{\sigma_1 \sigma_2}$. 
 
For CDD and QDD,  we have $Z_k = m_k^* + W_k, k=1, 2$, where $m_k^*$ is the value sent by the $k$-th sensor, i.e., $m_{k0}$ or $m_{k1}$, $k=1,2$. 
Therefore,  $(Z_1, Z_2|m_1^*, m_2^*) \sim G(m_1^*, m_2^*, \rho, \sigma_{c_1}, \sigma_{c_2})$.  $m_k^*$, $k=1,2$, are fixed values for CDD but must be optimally selected for QDD. 
At the FC, 
\begin{eqnarray}
f_{Z}(z_1,z_2|H_j) = \sum_{m_1^*}\sum_{m_2^*} G\left(m_1^*, m_2^*, \rho, \sigma_{c_1}, \sigma_{c_2} \right) \nonumber 
\\P(m_1^*|H_j)P(m_2^*|H_j), \quad j=1,2.\nonumber
\end{eqnarray}
LRT is then used at the FC to make a decision for received values of $z_1$ and $z_2$. 
 
\begin{figure}[htbp]
	\centerline{\includegraphics[width=.44\textwidth]{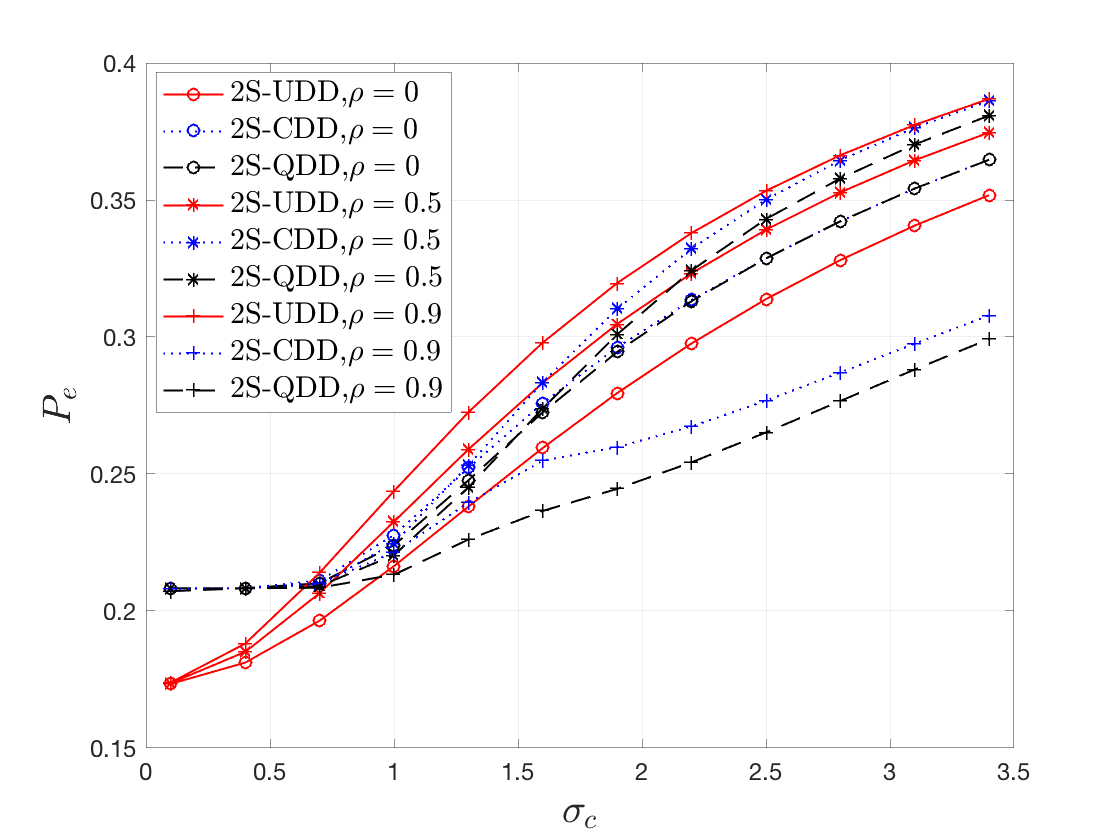}}
	\caption{Optimal sensor rules. $\pi_1=0.5, \mu=1, \sigma_s=1.5$, 2-sensors, $D=2$ }
	\label{fig:p1_05_rho_pe}
\end{figure}

Fig. \ref{fig:p1_05_rho_pe} shows the decision performance for correlated channels modeled by bivariate Gaussian. To check the effect of correlation, we set $\sigma_{s_1}=\sigma_{s_2}=1.5$ and $\sigma_{c_1}=\sigma_{c_2}=\sigma_c$, where $\sigma_c$ varies in a range.  $\rho=\{0, 0.5, 0.9\}$ are considered.  

From the figure, in the uncorrelated case, i.e., $\rho=0$, UDD always outperforms CDD and QDD, while QDD performs slightly better than or equal to CDD. When $\rho=0.5$, CDD and UDD start to outperform UDD in a certain range of $\sigma_c$, such as around $0.7<\sigma_c<1.6$ for CDD, and around $0.7<\sigma_c<2.2$ for QDD. When $\rho=0.9$, for $\sigma_c$ larger than a value around $\sigma_c=0.7$, QDD works much better than CDD, and both QDD and CDD have significant improvement over UDD.  It can be observed that when $\rho$ increases, the performance of UDD is always getting worse. This can be understood that for UDD, the 2-sensor case degenerates towards the 1-sensor case when $\rho \rightarrow 1$. However, an interesting observation for both CDD and QDD is that the Bayes' error increases for a moderate correlation coefficient, such as $\rho=0.5$, but decreases significantly when $\rho$ becomes very large, such as $\rho=0.9$. This phenomenon reflects the findings in \cite{kasasbeh17:noise, sun22:performance}. It can be inferred that when $\rho \rightarrow 1$, by checking $Z_1-Z_2$ at the FC, one can completely resolve which the two values of $m_1^*$ and $m_2^*$ are, if the differences between any two of $m_{10}, m_{11}, m_{20}, m_{21}$ are different. As a result, in such an extreme case of $\rho \rightarrow 1$, by optimally selecting values of $m_{10}, m_{11}, m_{20}$ and $m_{21}$, the reporting channel can be treated as error-free. For the CDD case, however, since $m_{10}=-m_{11}$ and $m_{20}=-m_{21}$, one cannot completely resolve $m_1^*$ and $m_2^*$ even if $\rho \rightarrow 1$. For example,  both $m_{10}-m_{20}$ and $m_{11}-m_{21}$ give the same value. Therefore, CDD cannot take the full advantage of the high correlation as QDD does.

\subsection{Asymptotic performance and Chernoff information}

Assuming i.i.d. AWGN for both sensing and reporting stages, deciding regions of $(\sigma_s, \sigma_c)$ where UDD and QDD outperforms each other is of great interest. However, this is a difficult task. For $D\geq 2$ and $N\geq 2$, the process by solving the optimization problem of $P_1$ seems insurmountable. It is possible, however, to readily compare UDD and QDD for the one-sensor case with $\pi_1=0.5$, and for the case of using asymptotically many sensors. 

For the one-sensor case with equal priors, we know that for QDD, the optimal sensor decision is $t=0, m_0=-m_1=-\sqrt{E_u}$,  and the Bayes' error is represented by (\ref{eqn:pe_1s}) where $E_u = \sigma_s^2+\mu^2$.  For UDD, it is a test of $\mathcal{N}(\mu, \sigma_s^2+\sigma_c^2)$ under $H_1$ against $\mathcal{N}(-\mu, \sigma_s^2+\sigma_c^2)$ under $H_0$ at the FC. With $\pi_1=0.5$, the Bayes' error of UDD is simply $Q( \frac{\mu} {\sqrt{\sigma_s^2+\sigma_c^2} } )$. 

For the asymptotic case of a large number of sensors, the decision error decreases exponentially with an exponent determined by the Chernoff information \cite{cover:elements}. Given two distributions $f_j(z) = p(z|H_j)$, $j=0, 1$, the Chernoff information is defined as 
\begin{eqnarray}
C(f_1, f_0) = D(f_\lambda||f_j), \qquad j=0, 1 
\end{eqnarray} 
where $D(f_\lambda||f_j)=E_{f_\lambda}\left[ \log \frac{f_\lambda}{f_j} \right]$ is the Kullback-Leibler divergence between the distributions of $f_\lambda$ and $f_j$, $j=0$ or $1$. $f_\lambda$ is a distribution in the form 
\begin{eqnarray}
f_\lambda(z) = \frac{f_1(z)^\lambda f_0(z)^{1-\lambda}}{\int f_1(v)^\lambda f_0(v)^{1-\lambda} dv}
\end{eqnarray} 
with $\lambda$ selected such that $D(f_\lambda||f_0) = D(f_\lambda|f_1)$. 

For the UDD case, at the FC,  $f_1 \sim \mathcal{N}(\mu, \sigma^2)$ and $f_0 \sim \mathcal{N}(-\mu, \sigma^2)$, where $\sigma^2=  \sigma_s^2+\sigma_c^2$. Therefore, $f_\lambda \sim \mathcal{N}(0,\sigma^2)$ and the Chernoff information is 
\begin{eqnarray}
C_u(f_1, f_0) = \frac{\mu^2}{2(\sigma_s^2+\sigma_c^2)}.
\end{eqnarray}

For the QDD case with $D=2$ and $t=0, m_1=-m_0=\sqrt{E_u}$, $f_1$ and $f_0$ at the FC are symmetric Gaussian mixtures. With $\lambda=1/2$, numerical method \cite{hershey07:approximating} can be applied to obtain the $C_q(f_1, f_0)$. Therefore, we can compare $C_u(f_1, f_0)$ and $C_q(f_1,f_0)$ on the plane of $(\sigma_s, \sigma_c)$ to observe the regions where one would be preferred to the other in the asymptotic case. 

\begin{figure}[htbp]
	\centerline{\includegraphics[width=.44\textwidth]{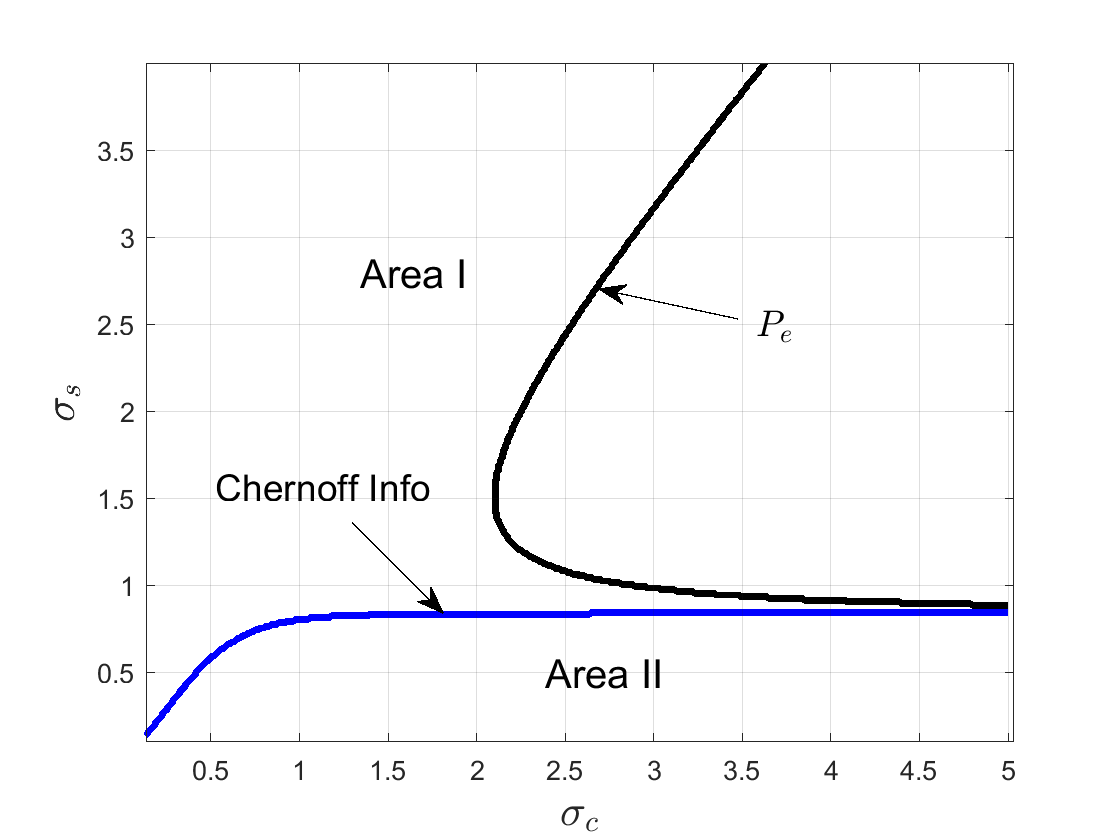}}
	\caption{$\pi_1=0.5, \mu=1$, regions where either UDD or QDD outperforms for cases with 1-sensor and asymptotically many sensors. $D=2$.}
	\label{fig:sig_sc}
\end{figure}

Fig. \ref{fig:sig_sc} shows the performance comparison of UDD and QDD for the scenarios of using 1-sensor and asymptotically many sensors, for different $(\sigma_s, \sigma_c)$ values. These curves are obtained by equating the Bayes' errors of UDD and QDD for the 1-sensor case, and by equating the Chernoff information of UDD and QDD for the asymptotic case. The curves show where $P_e$ values (or the Chernoff information) are equal for UDD and QDD when $D=2$.  Each curve divides the $(\sigma_s, \sigma_c)$ plane into two regions: one is the left (or bottom) to the curve where QDD outperforms, and its complement, i.e., right (or top) to the curve where UDD outperforms.  

Interesting observations are as follows.  
1) In the 1-sensor case, the effect of $\sigma_c$ seems more prominent. For $\sigma_c\leq 2.1$, QDD is always preferred to UDD, regardless of $\sigma_s$.  
2) For the asymptotic case, $\sigma_s$ seems to play a dominant role. For $\sigma_s \geq 0.85$, UDD always performs better than QDD, regardless of $\sigma_c$.

\subsection{Discussion and Future Work}

To solve $P_1$, the global optimization toolbox in Matlab\textsuperscript{\textregistered} is utilized. The problem is in general non-convex with multiple local optimal solutions. For example, in \ref{correlated} where $\pi_1=0.5$ and two sensors experience i.i.d. distortions in sensing and reporting, if a set of values $\{t_1, m_{10}, m_{11}, t_2, m_{20}, m_{21}\}$ is optimal, due to the symmetry, the sets of values $\{t_2, m_{20}, m_{21}, t_1, m_{10}, m_{11}\}$ and $\{ -t_1, -m_{11}, -m_{10}, -t_2, -m_{21}, -m_{20}\}$ are also optimal. In order to take the advantage of high correlation, $\{m_{10}, m_{11}, m_{20}, m_{21}\}$ are generally different for a high $\rho$ value. Therefore, to determine which region in terms of ${\sigma_s, \sigma_c}$ where the problem could be convex is an interesting topic, which should be helpful to develop more efficient and effective optimization solvers that are particularly important when the number of sensors increases. 

A long-run presumption held in CDD is that increasing $D$ leads to better decision performance at the FC or at least no worse. This presumption is unlikely correct when considering power constraint. As we can find when $D$ increases, QDD eventually becomes UDD. Therefore, for those scenarios when QDD with $D=2$ is better than UDD, increasing $D$ apparently will degrade the performance of QDD at certain points. Though this work considers $D=2$ only, the questions of which is the optimal $D$ for QDD, for a given number of sensors, and what is the limit of the performance gap between QDD and UDD are of fundamental interest. In addition, the sensor observations could be correlated as well \cite{willett00:good,kasasbeh17:noise}. In such a case, it is expected that LRQ will not be the optimal sensor rules or could be optimal only for certain situations that depend on the relationship among the levels of signals and noise, and the correlation coefficient. In-depth research is needed for the above issues.

\section{Conclusion}
In this paper, we present a new paradigm in distributed detection, i.e., Quantized-but-uncoded Distributed Detection (QDD). QDD is an extension of the quantized-and-Coded Distributed Detection (CDD), which can take into account the power constraint that is missed in CDD. We have proved the optimality of using LRQs as the sensor rules in QDD for $D=2$, when the sensor observations are independent. We also showed that for one-sensor case, the sensor decision rule is in general no longer ``channel blind". Through numerical results, we have compared the performance of UDD, CDD and QDD for certain scenarios where both sensing and reporting noise exist. This work serves as an attempt to answer the fundamental question in DD, i.e., ``Quantize or not quantize?" The work in this paper may open up an interesting territory that needs more research efforts to reveal the full picture of the preferred regions of UDD and QDD.

\bibliographystyle{IEEEtran}
\bibliography{math,DD_journal,DD_conf,Xingjian,Cao,Copulas}

%
%

\end{document}